\documentclass[sigconf]{acmart}

\usepackage{balance}
\usepackage{csquotes}

\def\BibTeX{{\rm B\kern-.05em{\sc i\kern-.025em b}\kern-.08emT\kern-.1667em\lower.7ex\hbox{E}\kern-.125emX}}
    
\copyrightyear{2020}
\acmYear{2020}
\setcopyright{acmlicensed}
\acmConference[AIES '20] {2020 AAAI/ACM Conference on AI, Ethics, and Society}{February 7--8, 2020}{New York, NY, USA}
\acmBooktitle{2020 AAAI/ACM Conference on AI, Ethics, and Society (AIES'20), February 7--8, 2020, New York, NY, USA}
\acmPrice{15.00}
\acmDOI{10.1145/3375627.3375855}
\acmISBN{978-1-4503-7110-0/20/02}

\begin{document}

\fancyhead{}

\title{Robot Rights? Let's Talk about Human Welfare Instead}



\author{Abeba Birhane}
\affiliation{%
 \institution{School of Computer Science\\University College Dublin}
 \city{Dublin}
 \country{Ireland}}
\email{abeba.birhane@ucdconnect.ie}

\author{Jelle van Dijk}
\affiliation{%
 \institution{Department of Design, Production and Management\\
 University of Twente }
 \city{Enschede}
 \country{Netherlands}}
\email{jelle.vandijk@utwente.nl}


%
\renewcommand{\shortauthors}{Trovato and Tobin, et al.}

%
\begin{abstract}
The `robot rights' debate, and its related question of `robot responsibility', invokes some of the most polarized positions in AI ethics. While some advocate for granting robots rights on a par with human beings, others, in a stark opposition argue that robots are not deserving of rights but are objects that should be our slaves. Grounded in post-Cartesian philosophical foundations, we argue not just to deny robots `rights', but to deny that robots, as artifacts emerging out of and mediating human being, are the kinds of things that could be granted rights in the first place. Once we see robots as mediators of human being, we can understand how the `robot rights' debate is focused on first world problems, at the expense of urgent ethical concerns, such as machine bias, machine elicited human labour exploitation, and erosion of privacy all impacting society's least privileged individuals. We conclude that, if human being is our starting point and human welfare is the primary concern, the negative impacts emerging from machinic systems, as well as the lack of taking responsibility by people designing, selling and deploying such machines, remains the most pressing ethical discussion in AI.
\end{abstract}

%
%

%
\keywords{Robot rights, AI ethics, embodiment, human welfare}

%

%
\maketitle

\section{The Debate: Robot Rights}
Ethicists have been discussing the notion of `robot rights': the idea that we should grant (future) artificially intelligent machines `rights', comparable to `human rights', courtesy of their constitution as intelligent, autonomous agents. Some promote robot rights within an overall techno-optimistic, materialistic worldview, arguing we must avoid any a priori `biological chauvinism'. The reasoning goes; if machines would bring forth the sort of agency that we attribute to ourselves, we have no reason not to grant them the sorts of rights we grant ourselves \cite{dennett1987intentional, brooks2000will, asaro2006should}.

A more critical, emancipatory strand of robot ethics claims that granting robots rights is not only ethically justified, but more fundamentally helps to reflect on existing undercurrents in (Western) ethical debates. Discussing robot rights helps to undo ethics of its implicit paternalistic, Western oppressive foundations and contribute to the emancipation of oppressed groups such as women and people of colour \cite{gunkel2018robot}.

In stark contrast, some claim we actually should call robots our slaves \cite{bryson2010robots}. Bryson, one of the advocates of this position, is well aware of the connotations implied by the term slave. She explains slavery historically means dehumanisation, something most cultures have since come to be opposed to, for very good reasons:

\begin{displayquote}
``Given the very obviously human beings that have been labelled inhuman in the global culture's very recent past, many seem to have grown wary of applying the label at all. For example, Dennett \cite{dennett1987intentional} argues that we should allocate the rights of agency to anything that appears to be best reasoned about as acting in an intentional manner. ...Dennett's ... generosity is almost definitionally nice.'' \cite[p.~2]{bryson2010robots}  
\end{displayquote}

Bryson however disagrees with Dennett. Granting robots rights, she reasons, is not always nice. Human well-being should be our prime concern and any concerns with robots should never distract us from the real target. We fully agree with her here. 

Yet we disagree robots should be treated as `slaves'. In defense of her position, Bryson states: ``But surely dehumanization is only wrong when it's applied to someone who really is human?'' Our position would be that `dehumanization' is not so much wrong for robots, it is impossible. One cannot dehumanize something that wasn't human to begin with. If one uses the term slave, one implicitly assumes that the being one so names is the kind of being can be `dehumanized'. One has already implicitly `humanized' the robot, before subsequently enslaving it. One should obviously not enslave someone first taken to be human. 

Bryson already accepts part of framing of the narrative of robo-ethics, where a discussion to consider the ontological status of robots in relation to rights is legitimate in principle. Our position is that the entire discussion is completely misguided. At best, robot ethics debates are First World philosophical musings, too disengaged from actual affairs of humans in the real world. In the worst case, it may contain bad faith --- the white, male academic's diminutive characterization of actually oppressed people and their fight for rights, by appealing to `reason'. 

\section{A Summary of our Argument}

Some may argue that the idea of robot rights is a peculiar, irrelevant discussion existing only at the fringes of AI ethics research more broadly construed, and as such devoting our time to it would not be paying justice to the important work done in that field. But the idea of robot rights is, in principle, perfectly legitimate if one stays true to the materialistic commitments of artificial intelligence: in principle it should be possible to build an artificially intelligent machine, and if we would succeed in doing so, there would be no reason not to grant this machine the rights we attribute to ourselves. Our critique therefore is not that the reasoning is invalid as such, but rather that we should question its underlying assumptions. Robot rights signal something more serious about AI technology, namely, that, grounded in their materialist techno-optimism, scientists and technologists are so preoccupied with the possible future of an imaginary machine, that they forget the very real, negative impact their intermediary creatures - the actual AI systems we have today - have on actual human beings. In other words: the discussion of robot rights is not to be separated from AI ethics, and AI ethics should concern itself with scrutinizing and reflecting deeply on underlying assumptions of scientists and engineers, rather than seeing its project as 'just' a practical matter of  discussing the ethical constraints and rules that should govern AI technologies in society.

Our starting point is not to deny robots `rights', but to deny that robots are the kinds of beings that could be granted or denied rights. We suggest it makes no sense to conceive of robots as slaves, since `slave' falls in the category of being that robots aren't. Human beings are such beings. We believe animals are such beings (though a discussion of animals lies beyond the scope of this paper). We take a post-Cartesian, phenomenological view in which being human means having a lived embodied experience, which itself is embedded in social practices. Technological artifacts form a crucial part of this being, yet artifacts themselves are not that same kind of being. The relation between human and technology is tightly intertwined, but not symmetrical. 

Based on this perspective we turn to the agenda for AI ethics. For some ethicists, to argue for robot rights, stems from their aversion against a human arrogance in face of the wider world. We too wish to fight human arrogance. But we see arrogance first and foremost in the techno-optimistic fantasies of the technology industry, making big promises to recreate ourselves out of silicon, surpassing ourselves with `super-AI' and `digitally uploading' our minds so as to achieve immortality, while at the same time exploiting human labour. Most debate on robot rights, we feel, is ultimately grounded in the same techno-arrogance. What we take from Bryson, is her plea to focus on the real issue: human oppression. We forefront the continual breaching of human welfare and especially of those disproportionally impacted by the development and ubiquitous integration of AI into society. Our ethical stance on human being is that being human means to interact with our surroundings in a respectful and just way. Technology should be designed to foster that. That, in turn, should be ethicists' primary concern. 

In what follows we first lay out our post-Cartesian perspective on human being and the role of technology within that perspective. Next, we explain why, even if robots should not be granted rights, we also reject the idea of the robot as a slave. In the final section, we call attention to human welfare instead. We discuss how AI, rather than being the potentially oppressed, is used as a tool by humans (with power) to oppress other humans, and how a discussion about robot rights diverts attention from the pressing ethical issues that matter. We end by reflecting on responsibilities, not of robots, but those of their human producers.
\subsection{A Post-Cartesian reframing}

The robot, like so many technologies created by humans, is created `in the image of ourselves'. But what is the self-image we use as a model? AI from its early days attempted to engineer a cognitivist interpretation of human thinking in the machine, which contains a (neo-)Cartesian distinction between, on the one hand the mental system, taken to be equivalent with the software of the machine, and the physical body, equivalent to the robot's physical parts. In contrast to Descartes' dualism however, cognitivists hold that the mental system is also physically realized, by mapping mental content onto physical processes (e.g., brain activation patterns). In general this is still the common sense conceptual model that underlies attempts at building intelligent machines. Consequently, for technologists and engineers a `human', on this model, can in principle be `built', because what it takes to be human, is ultimately a particular, complex configuration of physical processes \cite{churchland2013matter}. Starting from that model, the idea of robot rights makes perfect sense.

To understand how we reconceptualize the being of a robot, we need to look at our conception of human being, which rejects the image just described. In our post-Cartesian, phenomenologically inspired position, human being is a lived, embodied experience, or what Merleau-Ponty, following Husserl called, `being-in-the-world'. Embodied, enactive cognitive science, which follows this reasoning explains how biological living systems - living bodies - `enact' their perceptual world, through ongoing interactions with the environment \cite{di2018linguistic}. These interactions self-organise into sensorimotor couplings we may call habits or skills. Based on these couplings, we perceive (or rather `enact') things in the world in the first instance as affordances for action \cite{golonka2012gibson}. The things-as-affordances we perceive have direct relations with our bodily skills (Dreyfus \& Dreyfus, 2004). To give a common-sense example: a park bench `is' a different thing to a skateboarder, or a homeless person, than it is to a casual visitor. Embodied skills self-organize out of, and work to further sustain the organism. A second aspect concerns the inherently social nature of human being. We are always already situated within social practices, and the way we interact with and make sense of the world needs to be understood against this background. This view has been developed by the phenomenologists \cite{schutz1973structures}, and similarly developed through research on joint attention, situated practices \cite{lave1988cognition} and participatory sensemaking \cite{di2018linguistic}. 

Starting from human being as lived embodied interaction we can re-frame the role of technology. First, human-made artifacts attain their meaning as mediating our world enactment, by sustaining, breaking, changing, enriching sensorimotor couplings. This can be found in Heidegger's (1927) discussion of the hammer as being `ready-to-hand', and in Merleau-Ponty's (1962) discussion of the blind person's cane as extending the person's body. Within the more recent embodied cognitive science development, it relates to the idea of distributed cognition and the extended mind \cite{clark1998being}. Second, the meaning of artifacts must be understood within the context of our embedding social situation. In other words, things are what they are, because of the way they configure our social practices \cite{suchman2007human} and technology extends the biological body. Our conception of human being, then, is that we are and have always been fully embedded and enmeshed with our designed surroundings, and that we are critically dependent on this embeddedness for sustaining ourselves \cite{birhane2017descartes}.

The Cartesian illusion of setting ourselves apart from the natural, artificial and social world that we live in, spurred the project of building an artificial `intelligence', where intelligence is modeled on a human intelligence that is detached from the world and looks upon it, and the artifacts we create are things in that `objective', outside world. In contrast, Coeckelbergh's `social-relational' approach to machine ethics on the surface seems similar to our perspective \cite{coeckelbergh2010robot}. Yet he arrives at opposite conclusions. For Coeckelbergh, the `social-relational' describes the way people variably perceive artifacts, and perceiving them as `mere machines' is therefore just as valid as is perceiving them as `intelligent others'. In our view both Coeckelbergh and more traditional theorists all fail to realize how deeply embedded we already are with our technologies. A deep appreciation of this embeddedness does not entail artifacts should be seen as `agents like ourselves' (even if we socially talk about them that way): what we need to do is return to the realization that these technologies are always \textit{already part of ourselves}, as elements of our embodied being in the world\footnote{One may wonder if a perspective that builds on the technological mediation of lived experience should lead to the conclusion that `materiality has agency' (See Verbeek 2000 \cite{verbeek2000daadkracht}). If mediation by machines means those machines have agency, these should perhaps deserve rights. We reject the radical `symmetrical' position of Latour, in which objects and humans are networked as equals. Our position is more traditionally Heideggerian, in that we see technologies as building on and further sustaining (embodied, embedded, extended) human being. With Verbeek, however, we reject Heidegger's pessimistic dismissal of modern technologies: we think technologies can be recruited for the better, even if often used for the worst.} \cite{van2018designing}.

\subsection{Slaves are Humans Abused as Machines}
In \cite{gunkel2015rights} and elsewhere \cite{gunkel2018robot}, Gunkel builds a rhetoric in which he contrasts the ``seemingly cold and rather impersonal industrial robots'' to present-day social robots, which ``share physical and emotional spaces with the user'' \cite{gunkel2015rights}. ``For this reason'', he suggests ``it is reasonable to inquire about the social status and moral standing of these technologies'' (ibid). But we see no reason at all. Social robots are, as machines, as cold and impersonal as any machine. Or, looked at from another perspective, they are just as warm and personal as any machine, in the same way we can fall in love with a car, an espresso machine, or a house. None of this implies granting machines rights, at best it means we should take care of artifacts, as they were the product of hard labour, expressions of human creativity, received as a gift and so on. In other words: things configure social practices and taking care of things means taking care of ourselves. By taking care of things, we acknowledge their makers, we value their human designers, and we pay respect to a person that paid respect to us by presenting us a thing as a gift. 

Gunkel never falls into the trap of inventing fantasy futures with sentient machines to discuss robot rights. His issue is with the frame of mind that underlies opposition to robot rights, which in his view betrays an exclusionist `anthropocentric' reasoning, which not only marginalizes machines but has often been instrumental for excluding other human beings \cite{gunkel2015rights}. Citing \cite{stone1972should} he argues, ``Humans have defined numerous groups as less than human: slaves, woman, the 'other races,' children and foreigners...who have been defined as...as rightsless'' \cite[p~2]{gunkel2015rights}.	

But the very reason we judge the way slaves and women were (and still are) treated, as less than human, is that they are used as a means to an end, as `instruments' white men can use to get things done. The robot is the very model against which we judge whether humans are dehumanized. In Hannah Arendt's terminology: dehumanizing people means a reduction of their raison d'etre to mere labour, a mode of activity she distinguishes from `work' (a project), and `action' (political action) \cite{arendt1958human}. By putting actual slaves, women, and `other races' in one list with robots, one does not humanize them all, one \textit{dehumanizes} the actual humans in the list. Consider Coeckelbergh \cite{coeckelbergh2010robot}, when he writes: ``We have emancipated slaves, women, and some animals. First slaves and women were not treated as `men'. However, we made moral progress and now we consider them as human.'' This leads Coeckelbergh, to speculate on the equal emancipation of robots. But the choice of words suggests, even if unintended, a Western, white male's perspective on the matter (``\textit{we} emancipated women...''). The line of reasoning runs the risk of developing into: ``The women and slaves we liberated should not complain if we, enlightened men, decide to liberate some more!'' 

If our own reasoning is by contrast accused of as being `anthropocentric' then yes, this is exactly the point: robots are not humans, and our concern is with the welfare of human beings (see \cite{rushkoff2019team}).

\subsection{Robots are not Slaves}
As we said earlier, we disagree with treating the robots as slaves. We, while arguing against robot rights, use the (in)famous Milgram obedience to authority experiment to show why. 

We have to be aware of the difference between the way a person acts, and reflects back on their own actions, in a world they perceive to be actual, even if that world is in fact based on an illusion, versus the effect of a person's actions as seen from an outside observer's perspective. In the latter, `objective' frame, the participants in the Milgram experiment caused no harm, because the person who appeared to be screaming in pain was `in actuality', an actor. In the personal frame however, the world that the participant perceives to be real, they did do serious harm to another person - some even experienced having committed a murder. Being told in hindsight, that their experience was an illusion, did not help some of them to let go of that conclusion, and several were traumatized: 

\begin{displayquote}
``A New Haven Alderman complained to Yale authorities about the study: `I can't remember ever being quite so upset' (p. 132). One subject (\#716) checked mortality notices in the New Haven Register, for fear of having killed the learner. Another subject (\#501) was shaking so much he was not sure he would be able to drive home; according to his wife, on the way home he was shivering in the car and talked incessantly about his intense discomfort until midnight (p. 95). Subject 711 reported that `the experiment left such an effect on me that I spent the night in a cold sweat and nightmares because of fears that I might have killed that man in the chair' '' \cite[p~93]{brannigan2013stanley}. 
 \end{displayquote}
 
 If we look at the way we treat robots as through the eyes of a Milgram experiment participant, it would indeed be problematic to treat robots as slaves. The cultural-linguistic move of using the word slave, would mean - by analogy-  that in our enacted world, we would turn ourselves into slave owners, in the same `true' sense that the Milgram participants became murderers. 
 
 At the same time, the Milgram experiment frame also shows why we should object to the idea that the robot machine is treated unjustly. Following our Milgram logic, the robot is an actor. There is no real (third person objective) `recipient' of the unethical act. The only possible victim is the person who turned themselves into a slave owner, or, perhaps, society at large: if treating robots as slaves becomes commonplace, we may be engaging in social practices that we think are not making us better humans. Society may have reasons to reject such practices, even if no one would `do them for real' \cite{whitby2008sometimes}.
 
 But regardless of whether we think people are allowed to be `lured' into unethical acts with simulations, it remains the case that no injustice has been done to the actor that implemented the simulation, whether it is a human actor in Milgram's experiment, or a machine simulating a `sentient robot'. Perhaps the `robot as slave' can have a role in an educational setting, or as critical art, but there is no such thing as `robot rights', other than in fiction. 
 
\section{Let's Talk About Human Welfare Instead}
There are no robots that come close to the kind of `being' that humans are, and the kind of `being-with' that humans can have with other humans. Along with Hubert Dreyfus, we doubt if there ever will be \cite{dreyfus2004ethical}. Arguing for robot rights on the basis of future visions of sentient machines is speculative armchair philosophy at best. Meanwhile popular culture talks about actual AI and robots as if the intelligent machine is already there, while in fact, it is not. These sentiments betray the old cognitivist, Cartesian undercurrent in AI debates that sees the machines we create as `other agents, very much like ourselves', instead of what they are: mediators in embodied and socially situated human practices. 

One can maintain that it is romantic or ahistorical to think no technological progress could produce `true' AI in the future. But romanticism and lack of historical consciousness may be found on either side of the debate. Raymond Kurzweil \cite{kurzweil2005singularity}, for example, predicts that `mind uploading' will become possible by 2030s and sets the date for the singularity to occur by 2045. Romantic predictions like this, invariably envisioning breakthrough some decades into the future, have been recurring since the earliest days of digital technology, and all failed. It seems as if ``General AI'', ``the singularity'' and ``super-intelligence'' are for techno-optimists what doomsday is for religious cults. 

But it does not matter. Regardless of future predictions, what is of importance and urgency right now, is to call out the fact that farfetched romantic vistas of robot workers, robot care-givers and robot friends, and debating `the issue' of their supposed rights, contributes to real harm being done to individuals and groups, who are at present socioeconomically disadvantaged (which we elaborate in the next section). Whether or not our disbelief in the future existence of true AI will be proven wrong at some point, it is in any case less harmful than the recurring optimism about purely fictional futures. Because instead of steadily progressing towards a happy community of humans and `sentient AIs', techno-optimism contributes to the current development of dehumanizing technological infrastructure \cite{rushkoff2019team}. Debating the necessary conditions for robot rights keeps putting focus on (non-existent) machines, instead of on real people. In the next section we focus on what does exist: machines with software that we call `AI', which, in the reality of today, cause people harm.

\subsection{Robots are Used to Violate Human Rights}
Discussions of robot ethics, by portraying robots as intelligent systems as our primary concern, downplays the fact that we are currently amid artificially intelligent systems rapidly infiltrating every aspect of life. The real and urgent issues that are emerging with the mass deployment of seemingly invisible AI systems need to be discussed now because they currently impact large groups of people. 

The mass deployment of machines and AI today should propel us to examine commercial drives behind these machines as well as the harm and injustice the integration of machines into society brings. From perpetuation of historical and social bias and injustice \cite{benjamin2019race, eubanks2018automating, o2016weapons} to invasion of privacy \cite{zuboff2019age} to exploitation of human labour \cite{tubaro2019micro}, often for financial gains for private corporates, AI systems stand in opposition to human welfare. When AI systems are deployed and integrated into our day-to-day lives without critical examination and anticipation of emerging side-effects, they pose threats to human well-being. 

With the rise of machine learning, there is an increased appetite to hand much of our social, political and economical problems over to machines bringing with it corporate greed at the expense of human welfare and integrity \cite{zuboff2019age}. For the corporate world which produces a great proportion of current AI, profit marks its central objective, while for those deploying such technologies in various social sectors, AI seemingly provides a quick and cost-efficient solution to complex and messy social problems. However, the integration of these systems is proving to be a threat to people's welfare, integrity and privacy, especially those socioeconomically disadvantaged \cite{angwin2016machine, obermeyer2019dissecting, o2016weapons}. We discuss a number of these threats below. 

\subsection{Machine Bias and Discrimination}
It has become trivial to point out how decision-making processes in various social, political and economical sphere are assisted by automated systems. AI solutions pervade most spheres of life from screening potential employees to interviewing them, to predicting where criminal activity might occur (in some cases who might commit a crime) to diagnosing illnesses. These are highly contested and inherently political and moral issues that the technology industry is nonetheless treating as ``technical problems'' that can be quantified and automated.   

The automation of complex social, political and cultural issues requires that these complex, multivalent and contextual and continually moving concepts be quantified, measured, classified and captured through data \cite{mcquillan2018data}. Extrapolations, inferences and predictive models are then built often with real life actionable applications with grave consequences on society's most vulnerable. Machine learning systems that infer and predict individual behaviour and action, based on superficial extrapolations, are then deployed into the social world resulting in the emergence of various problems. These systems pick up social and historical stereotypes more than any deep fundamental causal explanations. In the process, individuals and groups, often at the margins of society that fail to fit stereotypical boxes suffer the undesirable consequences \cite{keyes2018misgendering}. A recurring theme within algorithmic bias, for example, shows that individuals and groups that have historically been marginalized are disproportionately impacted. This includes, for example, bias in detecting skin tones in pedestrians \cite{wilson2019predictive}; bias in predictive policing systems \cite{richardson2019dirty}; gender bias and discriminations in the display of STEM career ads \cite{lambrecht2019algorithmic}; racial bias in recidivism algorithms \cite{angwin2016machine}; bias in the politics of search engines \cite{introna2000defining}; bias and discrimination in medicine \cite{ferryman2018fairness,obermeyer2019dissecting}. 

AI, far from a future phenomenon waiting to happen, is here operating ubiquitously and with a disastrous impact on socially and historically marginalized groups. As Weiser remarks: ``The most profound technologies are those that disappear. They weave themselves into the fabric of everyday life until they are indistinguishable from it.''  Ubiquitous AI is inextricably intertwined with what it means to be a human being \cite{cheney2018we}. Yet the question is, how to best frame this intertwining conceptually? The typical narrative seems to conceive of AI technologies as some type of social partner that we will communicate and live with in ways comparable to the ways other human beings are bound up with our lives. In reality, no robot today is anywhere near that future vision. The actual situation we have today shows machine learning algorithms as embedded in seemingly mundane tools, supporting everyday tasks. These algorithms are influencing our basic 'being in the world' - the way we perceive and categorise the world, the agency we ourselves have in acting on it, in a more invisible, Weiserian sense, which makes it all the more insidious. 
For example, the humanoid robot known as Sophia, epitomizes an image that sits well with widely held conception of ``intelligent robots'' but whereas in fact, it has rudimentary engine and capabilities in reality. In comparison, iRobot's Roomba, while portrayed as a harmless household machine, exerts much more impact on our lives, and the dark side of it, is that it serves as a surveillance tool that continually harvests data about our homes. It is easy to overlook the dangers that the Roomba pauses to our privacy as the machine fades into the back-ground and becomes silently incorporated into our day-to-day life. iRobot's Roomba ``autonomous'' vacuum cleaner is fitted with a camera, sensors and software enabling it to build maps of the private sanctuary of our home, while tracking its own location\cite{zuboff2019age}. In combination with other IoT devices, the Roomba can be used to supposedly map our habits,behavours, activities.

Most AI companies boost on capabilities to be able to provide insights into the human psyche. Financial interests of companies and engineers that collect, evaluate data and algorithmically interpret and predict behaviors drive AI research and development. As such ``smart'' systems infiltrate day-to-day life from the IoT devices to ``smart home'' all designed to render all corners of lived experience as behavioral data \cite{zuboff2019age}. Envisioning a future human-like intelligent system while putting aside such ubiquitous and invasive systems which are a thereat to privacy and human welfare, shows misplaced concern, to say the least. The integration of machinic systems into social and human affairs poses immediate danger, especially to disfranchised people that need the most protection (O'Neil, 2016). Taking ethical concerns seriously means, we argue, prioritizing welfare of people, especially those often disproportionally impacted by the integration of machinic systems into daily life.  

\subsection{Looking Under the AI Hood: Human labour}
If we look at robot rights taking real, existing technologies and the human practices that they mediate as a starting point, we realize that it is inherently difficult to draw a boundary around the (artificial) entity that would need to be granted rights. In fact, attempts to look at what constitutes current intelligent and seemingly autonomous systems reveals that far from being fully autonomous, these systems function on exploitive human labour. From robots to `autonomous' vehicles to sophisticated image recognition systems, all machines rely heavily on human input. Systems that are perceived as `autonomous' are never fully autonomous but instead human-machine systems. 

Furthermore, as Bainbridge \cite{bainbridge1983ironies} remarks ``the more advanced the system is, the more crucial the contribution of the human.'' This still remains the case for current intelligent systems \cite{baxter2012ironies, strauch2017ironies}. ``The more we depend on technology and push it to its limits, the more we need highly-skilled, well-trained, well-practiced people to make systems resilient, acting as the last line of defence against the failures that will inevitably occur.'' \cite{baxter2012ironies}. AI systems rely not only on high-skilled and well-paid engineers and scientists but also are heavily dependent on the contribution of the less visible and low-paid human labour, referred to as ``microwork'' or ``crowd work''. From annotating and adding labels to images, to identifying objects in a photograph, to sorting items on a list, these low paid crowd works prepare ``training'' data for machines \cite{tubaro2019micro}. As well as poorly paid work, unpaid human labour fuels the development of proprietary intelligent systems where private corporates control and benefit from. Google's reCAPTCHA, which first emerged as a technique to prevent spam, then used to digitize old books, and later as means to availing training data for machine learning systems such as `autonomous cars' and face recognition software \footnote{see Schmieg \& Lorusso (2017) Five Years of Captured Captchas. \url{http://five.yearsofcapturedcapt.ch/as}} is one such example. AI thrives on the backbone of human labour and as Bainbridge \cite{bainbridge1983ironies} remarked in Ironies of Automation, the more advanced the technology, the more crucial the contribution of the human. As image recognition systems become more advanced, the images that humans have to label and annotate become harder, making the task more difficult for people.

What a close examination of the workings of intelligent systems reveals is that, not only are AI systems always human-machine systems but they are also inseparable from the profit driven business models of the industry that develop and deploy them. AI systems are intermeshed with humans (not separate entities) and serve as a constitutive influence of our being. Using humans to do low-paid micro-work to make AI possible is, in our view, dehumanizing, following Hanna Arendts' category of labour. More generally, the power imbalance between those that produce and control technology and the prioritization of financial profits as central objectives means that machines are used by the powerful and privileged as tools that hamper human welfare. 

\subsection{In Conclusion: Taking Back Control}
In October 2019, Emily Ackerman, a wheelchair user, described her experience of being ``trapped'' on the road by a Starship Technologies robot. These robots use the curb ramp to cross streets and one blocked her access to the sidewalk. ``I can tell, as long as they [robots] continue to operate, they are going to be a major accessibility and safety issue'', complains Ackerman \footnote{Pitt pauses testing of Starship robots due to safety concerns | The PittNews. Wolfe, E. (2019) \url{https://pittnews.com/article/151679/news/pitt-pauses-testing-of-starship-robots-due-to-safety-concerns/}}. Questions such as do these robots have the right to use public space and whether a ban might infringe `their' rights, as debated within the `robot rights' discourse, prioritize the wrong concerns. It is like protecting the gun instead of the victim. Primary concern should be with the welfare of marginalized groups (wheelchair users, in this case) which are disproportionally impacted by the integration of technology into our everyday lifeworlds. 

When a philosopher is contemplating what would be the ontological conditions for anything to be granted rights, it is easy to end up in arguments that compare `the rights of the human' with `the rights of the robot'. But this comparison is based on the, in our view, false belief that sees human being as just a complicated machine, and in thinking that complicated human-made machines could therefore replicate human being. Based on the post-Cartesian embodied perspective we hold that while human being may incorporate, and extend itself in creating and using machines, the intelligent machine remains a fantasy idea. What is more, what we see is that in pursuit of this fantasy, real machines are created, and these very real, data processing pattern recognition algorithms are increasingly \textit{getting in the way} of human well-being, up to the point of contributing to the dehumanization of real humans. 

Putting our feet back in reality, what we actually have at hand are situations in which a human being (a wheelchair user) is denied free movement by a machine, used by a corporate company who monopolizes public space for financial gain.

In closing, we turn to responsibility. In our view it is companies, engineers, policy makers, and the public at large, who are responsible to ensure the rights of individual people. One of the pressing issues in this day and age is that `intelligent' machines are increasingly used in sustaining forms of oppression. We do not `blame' the machines (they can take no blame), nor do we say machines must bear `responsibility' \cite{coeckelbergh2019artificial}, precisely because this would relieve those actually responsible from their duties. We agree that, in the complex networked society of today, it can be very complex if not often impossible to trace back accountability to individual people \cite{coeckelbergh2019artificial}. But this fact of life (it is complex) is no argument at all for making machines responsible. By making robots block the part of the pavement, a pavement that was designed to allow wheelchair uses to independently navigate city traffic, we take away part of the socio-technical embedding that supported a marginalized group in exerting their autonomy all for a business driven by financial gains.

More generally speaking, transferring ever more control over complex processes to intelligent machines - outsourcing our thinking and decision making, so to speak, to these technologies, may actually work against the empowerment of individual human beings, may even prevent them from taking the responsibilities we would expect to go together with having human rights.
%
\begin{acks}
This work is supported, in part, by Science Foundation Ireland grant 13/RC/2094.
\end{acks}

%
\bibliographystyle{ACM-Reference-Format}
\bibliography{acmart}


\begin{thebibliography}{43}


\ifx \showCODEN    \undefined \def \showCODEN     #1{\unskip}     \fi
\ifx \showDOI      \undefined \def \showDOI       #1{#1}\fi
\ifx \showISBNx    \undefined \def \showISBNx     #1{\unskip}     \fi
\ifx \showISBNxiii \undefined \def \showISBNxiii  #1{\unskip}     \fi
\ifx \showISSN     \undefined \def \showISSN      #1{\unskip}     \fi
\ifx \showLCCN     \undefined \def \showLCCN      #1{\unskip}     \fi
\ifx \shownote     \undefined \def \shownote      #1{#1}          \fi
\ifx \showarticletitle \undefined \def \showarticletitle #1{#1}   \fi
\ifx \showURL      \undefined \def \showURL       {\relax}        \fi
\providecommand\bibfield[2]{#2}
\providecommand\bibinfo[2]{#2}
\providecommand\natexlab[1]{#1}
\providecommand\showeprint[2][]{arXiv:#2}

\bibitem[\protect\citeauthoryear{Angwin, Larson, Mattu, and Kirchner}{Angwin
  et~al\mbox{.}}{2016}]%
        {angwin2016machine}
\bibfield{author}{\bibinfo{person}{Julia Angwin}, \bibinfo{person}{Jeff
  Larson}, \bibinfo{person}{Surya Mattu}, {and} \bibinfo{person}{Lauren
  Kirchner}.} \bibinfo{year}{2016}\natexlab{}.
\newblock \showarticletitle{Machine bias}.
\newblock \bibinfo{journal}{\emph{ProPublica, May}}  \bibinfo{volume}{23}
  (\bibinfo{year}{2016}), \bibinfo{pages}{2016}.
\newblock


\bibitem[\protect\citeauthoryear{Arendt}{Arendt}{1958}]%
        {arendt1958human}
\bibfield{author}{\bibinfo{person}{Hannah Arendt}.}
  \bibinfo{year}{1958}\natexlab{}.
\newblock \showarticletitle{The human condition. Chicago and London}.
\newblock \bibinfo{journal}{\emph{The University of Chicago Press. Available at
  http://www. kenvale. edu.
  au/Text/1313732697393-6317/uploadedFiles/1313731768783-9269. pdf. Accessed on
  July}}  \bibinfo{volume}{8} (\bibinfo{year}{1958}), \bibinfo{pages}{2012}.
\newblock


\bibitem[\protect\citeauthoryear{Asaro}{Asaro}{2006}]%
        {asaro2006should}
\bibfield{author}{\bibinfo{person}{Peter~M Asaro}.}
  \bibinfo{year}{2006}\natexlab{}.
\newblock \showarticletitle{What should we want from a robot ethic}.
\newblock \bibinfo{journal}{\emph{International Review of Information Ethics}}
  \bibinfo{volume}{6}, \bibinfo{number}{12} (\bibinfo{year}{2006}),
  \bibinfo{pages}{9--16}.
\newblock


\bibitem[\protect\citeauthoryear{Bainbridge}{Bainbridge}{1983}]%
        {bainbridge1983ironies}
\bibfield{author}{\bibinfo{person}{Lisanne Bainbridge}.}
  \bibinfo{year}{1983}\natexlab{}.
\newblock \showarticletitle{Ironies of automation}.
\newblock In \bibinfo{booktitle}{\emph{Analysis, design and evaluation of
  man--machine systems}}. \bibinfo{publisher}{Elsevier},
  \bibinfo{pages}{129--135}.
\newblock


\bibitem[\protect\citeauthoryear{Baxter, Rooksby, Wang, and
  Khajeh-Hosseini}{Baxter et~al\mbox{.}}{2012}]%
        {baxter2012ironies}
\bibfield{author}{\bibinfo{person}{Gordon~D Baxter}, \bibinfo{person}{John
  Rooksby}, \bibinfo{person}{Yuanzhi Wang}, {and} \bibinfo{person}{Ali
  Khajeh-Hosseini}.} \bibinfo{year}{2012}\natexlab{}.
\newblock \showarticletitle{The ironies of automation: still going strong at
  30?}. In \bibinfo{booktitle}{\emph{ECCE}}. \bibinfo{pages}{65--71}.
\newblock


\bibitem[\protect\citeauthoryear{Benjamin}{Benjamin}{2019}]%
        {benjamin2019race}
\bibfield{author}{\bibinfo{person}{Ruha Benjamin}.}
  \bibinfo{year}{2019}\natexlab{}.
\newblock \bibinfo{booktitle}{\emph{Race after technology: Abolitionist tools
  for the new jim code}}.
\newblock \bibinfo{publisher}{John Wiley \& Sons}.
\newblock


\bibitem[\protect\citeauthoryear{Birhane}{Birhane}{2017}]%
        {birhane2017descartes}
\bibfield{author}{\bibinfo{person}{Abeba Birhane}.}
  \bibinfo{year}{2017}\natexlab{}.
\newblock \showarticletitle{Descartes Was Wrong:‘A Person Is a Person through
  Other Persons’}.
\newblock \bibinfo{journal}{\emph{Aeon}} (\bibinfo{year}{2017}).
\newblock


\bibitem[\protect\citeauthoryear{Brannigan}{Brannigan}{2013}]%
        {brannigan2013stanley}
\bibfield{author}{\bibinfo{person}{Augustine Brannigan}.}
  \bibinfo{year}{2013}\natexlab{}.
\newblock \showarticletitle{Stanley Milgram’s obedience experiments: A report
  card 50 years later}.
\newblock \bibinfo{journal}{\emph{Society}} \bibinfo{volume}{50},
  \bibinfo{number}{6} (\bibinfo{year}{2013}), \bibinfo{pages}{623--628}.
\newblock


\bibitem[\protect\citeauthoryear{Brooks}{Brooks}{2000}]%
        {brooks2000will}
\bibfield{author}{\bibinfo{person}{Rodney Brooks}.}
  \bibinfo{year}{2000}\natexlab{}.
\newblock \showarticletitle{WILL ROBOTS DEMAND EQUAL RIGHTS?}
\newblock \bibinfo{journal}{\emph{TIME-NEW YORK-}} \bibinfo{volume}{155},
  \bibinfo{number}{25} (\bibinfo{year}{2000}), \bibinfo{pages}{86--86}.
\newblock


\bibitem[\protect\citeauthoryear{Bryson}{Bryson}{2010}]%
        {bryson2010robots}
\bibfield{author}{\bibinfo{person}{Joanna~J Bryson}.}
  \bibinfo{year}{2010}\natexlab{}.
\newblock \showarticletitle{Robots should be slaves}.
\newblock \bibinfo{journal}{\emph{Close Engagements with Artificial Companions:
  Key social, psychological, ethical and design issues}}
  (\bibinfo{year}{2010}), \bibinfo{pages}{63--74}.
\newblock


\bibitem[\protect\citeauthoryear{Cheney-Lippold}{Cheney-Lippold}{2018}]%
        {cheney2018we}
\bibfield{author}{\bibinfo{person}{John Cheney-Lippold}.}
  \bibinfo{year}{2018}\natexlab{}.
\newblock \bibinfo{booktitle}{\emph{We are data: Algorithms and the making of
  our digital selves}}.
\newblock \bibinfo{publisher}{NYU Press}.
\newblock


\bibitem[\protect\citeauthoryear{Churchland}{Churchland}{2013}]%
        {churchland2013matter}
\bibfield{author}{\bibinfo{person}{Paul~M Churchland}.}
  \bibinfo{year}{2013}\natexlab{}.
\newblock \bibinfo{booktitle}{\emph{Matter and consciousness}}.
\newblock \bibinfo{publisher}{MIT press}.
\newblock


\bibitem[\protect\citeauthoryear{Clark}{Clark}{1998}]%
        {clark1998being}
\bibfield{author}{\bibinfo{person}{Andy Clark}.}
  \bibinfo{year}{1998}\natexlab{}.
\newblock \bibinfo{booktitle}{\emph{Being there: Putting brain, body, and world
  together again}}.
\newblock \bibinfo{publisher}{MIT press}.
\newblock


\bibitem[\protect\citeauthoryear{Coeckelbergh}{Coeckelbergh}{2010}]%
        {coeckelbergh2010robot}
\bibfield{author}{\bibinfo{person}{Mark Coeckelbergh}.}
  \bibinfo{year}{2010}\natexlab{}.
\newblock \showarticletitle{Robot rights? Towards a social-relational
  justification of moral consideration}.
\newblock \bibinfo{journal}{\emph{Ethics and Information Technology}}
  \bibinfo{volume}{12}, \bibinfo{number}{3} (\bibinfo{year}{2010}),
  \bibinfo{pages}{209--221}.
\newblock


\bibitem[\protect\citeauthoryear{Coeckelbergh}{Coeckelbergh}{2019}]%
        {coeckelbergh2019artificial}
\bibfield{author}{\bibinfo{person}{Mark Coeckelbergh}.}
  \bibinfo{year}{2019}\natexlab{}.
\newblock \showarticletitle{Artificial Intelligence, Responsibility
  Attribution, and a Relational Justification of Explainability}.
\newblock \bibinfo{journal}{\emph{Science and engineering ethics}}
  (\bibinfo{year}{2019}), \bibinfo{pages}{1--18}.
\newblock


\bibitem[\protect\citeauthoryear{Dennett}{Dennett}{1987}]%
        {dennett1987intentional}
\bibfield{author}{\bibinfo{person}{Daniel~C Dennett}.}
  \bibinfo{year}{1987}\natexlab{}.
\newblock \showarticletitle{The intentional stance. 1987}.
\newblock \bibinfo{journal}{\emph{Cambridge, MA}}  \bibinfo{volume}{802}
  (\bibinfo{year}{1987}).
\newblock


\bibitem[\protect\citeauthoryear{Di~Paolo, Cuffari, and De~Jaegher}{Di~Paolo
  et~al\mbox{.}}{2018}]%
        {di2018linguistic}
\bibfield{author}{\bibinfo{person}{Ezequiel~A Di~Paolo},
  \bibinfo{person}{Elena~Clare Cuffari}, {and} \bibinfo{person}{Hanne
  De~Jaegher}.} \bibinfo{year}{2018}\natexlab{}.
\newblock \bibinfo{booktitle}{\emph{Linguistic bodies: The continuity between
  life and language}}.
\newblock \bibinfo{publisher}{MIT Press}.
\newblock


\bibitem[\protect\citeauthoryear{Dreyfus and Dreyfus}{Dreyfus and
  Dreyfus}{2004}]%
        {dreyfus2004ethical}
\bibfield{author}{\bibinfo{person}{Hubert~L Dreyfus} {and}
  \bibinfo{person}{Stuart~E Dreyfus}.} \bibinfo{year}{2004}\natexlab{}.
\newblock \showarticletitle{The ethical implications of the five-stage
  skill-acquisition model}.
\newblock \bibinfo{journal}{\emph{Bulletin of Science, Technology \& Society}}
  \bibinfo{volume}{24}, \bibinfo{number}{3} (\bibinfo{year}{2004}),
  \bibinfo{pages}{251--264}.
\newblock


\bibitem[\protect\citeauthoryear{Eubanks}{Eubanks}{2018}]%
        {eubanks2018automating}
\bibfield{author}{\bibinfo{person}{Virginia Eubanks}.}
  \bibinfo{year}{2018}\natexlab{}.
\newblock \bibinfo{booktitle}{\emph{Automating inequality: How high-tech tools
  profile, police, and punish the poor}}.
\newblock \bibinfo{publisher}{St. Martin's Press}.
\newblock


\bibitem[\protect\citeauthoryear{Ferryman and Pitcan}{Ferryman and
  Pitcan}{2018}]%
        {ferryman2018fairness}
\bibfield{author}{\bibinfo{person}{Kadija Ferryman} {and}
  \bibinfo{person}{Mikaela Pitcan}.} \bibinfo{year}{2018}\natexlab{}.
\newblock \showarticletitle{Fairness in precision medicine}.
\newblock \bibinfo{journal}{\emph{Data \& Society}} (\bibinfo{year}{2018}).
\newblock


\bibitem[\protect\citeauthoryear{Golonka and Wilson}{Golonka and
  Wilson}{2012}]%
        {golonka2012gibson}
\bibfield{author}{\bibinfo{person}{Sabrina Golonka} {and}
  \bibinfo{person}{Andrew~D Wilson}.} \bibinfo{year}{2012}\natexlab{}.
\newblock \showarticletitle{Gibson’s ecological approach}.
\newblock \bibinfo{journal}{\emph{Avant: Trends in Interdisciplinary Studies 3
  (2)}} (\bibinfo{year}{2012}), \bibinfo{pages}{40--53}.
\newblock


\bibitem[\protect\citeauthoryear{Gunkel}{Gunkel}{2015}]%
        {gunkel2015rights}
\bibfield{author}{\bibinfo{person}{David~J Gunkel}.}
  \bibinfo{year}{2015}\natexlab{}.
\newblock \showarticletitle{The rights of machines: Caring for robotic
  care-givers}.
\newblock In \bibinfo{booktitle}{\emph{Machine Medical Ethics}}.
  \bibinfo{publisher}{Springer}, \bibinfo{pages}{151--166}.
\newblock


\bibitem[\protect\citeauthoryear{Gunkel}{Gunkel}{2018}]%
        {gunkel2018robot}
\bibfield{author}{\bibinfo{person}{David~J Gunkel}.}
  \bibinfo{year}{2018}\natexlab{}.
\newblock \bibinfo{booktitle}{\emph{Robot rights}}.
\newblock \bibinfo{publisher}{MIT Press}.
\newblock


\bibitem[\protect\citeauthoryear{Introna and Nissenbaum}{Introna and
  Nissenbaum}{2000}]%
        {introna2000defining}
\bibfield{author}{\bibinfo{person}{Lucas Introna} {and} \bibinfo{person}{Helen
  Nissenbaum}.} \bibinfo{year}{2000}\natexlab{}.
\newblock \showarticletitle{Defining the web: The politics of search engines}.
\newblock \bibinfo{journal}{\emph{Computer}} \bibinfo{volume}{33},
  \bibinfo{number}{1} (\bibinfo{year}{2000}), \bibinfo{pages}{54--62}.
\newblock


\bibitem[\protect\citeauthoryear{Keyes}{Keyes}{2018}]%
        {keyes2018misgendering}
\bibfield{author}{\bibinfo{person}{Os Keyes}.} \bibinfo{year}{2018}\natexlab{}.
\newblock \showarticletitle{The misgendering machines: Trans/HCI implications
  of automatic gender recognition}.
\newblock \bibinfo{journal}{\emph{Proceedings of the ACM on Human-Computer
  Interaction}} \bibinfo{volume}{2}, \bibinfo{number}{CSCW}
  (\bibinfo{year}{2018}), \bibinfo{pages}{88}.
\newblock


\bibitem[\protect\citeauthoryear{Kurzweil}{Kurzweil}{2005}]%
        {kurzweil2005singularity}
\bibfield{author}{\bibinfo{person}{Ray Kurzweil}.}
  \bibinfo{year}{2005}\natexlab{}.
\newblock \bibinfo{booktitle}{\emph{The singularity is near: When humans
  transcend biology}}.
\newblock \bibinfo{publisher}{Penguin}.
\newblock


\bibitem[\protect\citeauthoryear{Lambrecht and Tucker}{Lambrecht and
  Tucker}{2019}]%
        {lambrecht2019algorithmic}
\bibfield{author}{\bibinfo{person}{Anja Lambrecht} {and}
  \bibinfo{person}{Catherine Tucker}.} \bibinfo{year}{2019}\natexlab{}.
\newblock \showarticletitle{Algorithmic Bias? An Empirical Study of Apparent
  Gender-Based Discrimination in the Display of STEM Career Ads}.
\newblock \bibinfo{journal}{\emph{Management Science}} (\bibinfo{year}{2019}).
\newblock


\bibitem[\protect\citeauthoryear{Lave}{Lave}{1988}]%
        {lave1988cognition}
\bibfield{author}{\bibinfo{person}{Jean Lave}.}
  \bibinfo{year}{1988}\natexlab{}.
\newblock \bibinfo{booktitle}{\emph{Cognition in practice: Mind, mathematics
  and culture in everyday life}}.
\newblock \bibinfo{publisher}{Cambridge University Press}.
\newblock


\bibitem[\protect\citeauthoryear{McQuillan}{McQuillan}{2018}]%
        {mcquillan2018data}
\bibfield{author}{\bibinfo{person}{Dan McQuillan}.}
  \bibinfo{year}{2018}\natexlab{}.
\newblock \showarticletitle{Data science as machinic neoplatonism}.
\newblock \bibinfo{journal}{\emph{Philosophy and Technology}}
  \bibinfo{volume}{31}, \bibinfo{number}{2} (\bibinfo{year}{2018}),
  \bibinfo{pages}{253--272}.
\newblock


\bibitem[\protect\citeauthoryear{Obermeyer and Mullainathan}{Obermeyer and
  Mullainathan}{2019}]%
        {obermeyer2019dissecting}
\bibfield{author}{\bibinfo{person}{Ziad Obermeyer} {and}
  \bibinfo{person}{Sendhil Mullainathan}.} \bibinfo{year}{2019}\natexlab{}.
\newblock \showarticletitle{Dissecting Racial Bias in an Algorithm that Guides
  Health Decisions for 70 Million People}. In
  \bibinfo{booktitle}{\emph{Proceedings of the Conference on Fairness,
  Accountability, and Transparency}}. ACM, \bibinfo{pages}{89--89}.
\newblock


\bibitem[\protect\citeauthoryear{O'neil}{O'neil}{2016}]%
        {o2016weapons}
\bibfield{author}{\bibinfo{person}{Cathy O'neil}.}
  \bibinfo{year}{2016}\natexlab{}.
\newblock \bibinfo{booktitle}{\emph{Weapons of math destruction: How big data
  increases inequality and threatens democracy}}.
\newblock \bibinfo{publisher}{Broadway Books}.
\newblock


\bibitem[\protect\citeauthoryear{Richardson, Schultz, and Crawford}{Richardson
  et~al\mbox{.}}{2019}]%
        {richardson2019dirty}
\bibfield{author}{\bibinfo{person}{Rashida Richardson}, \bibinfo{person}{Jason
  Schultz}, {and} \bibinfo{person}{Kate Crawford}.}
  \bibinfo{year}{2019}\natexlab{}.
\newblock \showarticletitle{Dirty Data, Bad Predictions: How Civil Rights
  Violations Impact Police Data, Predictive Policing Systems, and Justice}.
\newblock \bibinfo{journal}{\emph{New York University Law Review Online,
  Forthcoming}} (\bibinfo{year}{2019}).
\newblock


\bibitem[\protect\citeauthoryear{Rushkoff}{Rushkoff}{2019}]%
        {rushkoff2019team}
\bibfield{author}{\bibinfo{person}{Douglas Rushkoff}.}
  \bibinfo{year}{2019}\natexlab{}.
\newblock \bibinfo{booktitle}{\emph{Team Human}}.
\newblock \bibinfo{publisher}{WW Norton and Company}.
\newblock


\bibitem[\protect\citeauthoryear{Schutz and Luckmann}{Schutz and
  Luckmann}{1973}]%
        {schutz1973structures}
\bibfield{author}{\bibinfo{person}{Alfred Schutz} {and} \bibinfo{person}{Thomas
  Luckmann}.} \bibinfo{year}{1973}\natexlab{}.
\newblock \bibinfo{booktitle}{\emph{The structures of the life-world}}.
  Vol.~\bibinfo{volume}{1}.
\newblock \bibinfo{publisher}{northwestern university press}.
\newblock


\bibitem[\protect\citeauthoryear{Stone}{Stone}{1972}]%
        {stone1972should}
\bibfield{author}{\bibinfo{person}{Christopher~D Stone}.}
  \bibinfo{year}{1972}\natexlab{}.
\newblock \showarticletitle{Should Trees Have Standing--Toward Legal Rights for
  Natural Objects}.
\newblock \bibinfo{journal}{\emph{S. CAl. l. rev.}}  \bibinfo{volume}{45}
  (\bibinfo{year}{1972}), \bibinfo{pages}{450}.
\newblock


\bibitem[\protect\citeauthoryear{Strauch}{Strauch}{2017}]%
        {strauch2017ironies}
\bibfield{author}{\bibinfo{person}{Barry Strauch}.}
  \bibinfo{year}{2017}\natexlab{}.
\newblock \showarticletitle{Ironies of automation: Still unresolved after all
  these years}.
\newblock \bibinfo{journal}{\emph{IEEE Transactions on Human-Machine Systems}}
  \bibinfo{volume}{48}, \bibinfo{number}{5} (\bibinfo{year}{2017}),
  \bibinfo{pages}{419--433}.
\newblock


\bibitem[\protect\citeauthoryear{Suchman and Suchman}{Suchman and
  Suchman}{2007}]%
        {suchman2007human}
\bibfield{author}{\bibinfo{person}{Lucy Suchman} {and} \bibinfo{person}{Lucy~A
  Suchman}.} \bibinfo{year}{2007}\natexlab{}.
\newblock \bibinfo{booktitle}{\emph{Human-machine reconfigurations: Plans and
  situated actions}}.
\newblock \bibinfo{publisher}{Cambridge university press}.
\newblock


\bibitem[\protect\citeauthoryear{Tubaro and Casilli}{Tubaro and
  Casilli}{2019}]%
        {tubaro2019micro}
\bibfield{author}{\bibinfo{person}{Paola Tubaro} {and}
  \bibinfo{person}{Antonio~A Casilli}.} \bibinfo{year}{2019}\natexlab{}.
\newblock \showarticletitle{Micro-work, artificial intelligence and the
  automotive industry}.
\newblock \bibinfo{journal}{\emph{Journal of Industrial and Business
  Economics}} (\bibinfo{year}{2019}), \bibinfo{pages}{1--13}.
\newblock


\bibitem[\protect\citeauthoryear{Van~Dijk}{Van~Dijk}{2018}]%
        {van2018designing}
\bibfield{author}{\bibinfo{person}{Jelle Van~Dijk}.}
  \bibinfo{year}{2018}\natexlab{}.
\newblock \showarticletitle{Designing for embodied being-in-the-world: A
  critical analysis of the concept of embodiment in the design of hybrids}.
\newblock \bibinfo{journal}{\emph{Multimodal Technologies and Interaction}}
  \bibinfo{volume}{2}, \bibinfo{number}{1} (\bibinfo{year}{2018}),
  \bibinfo{pages}{7}.
\newblock


\bibitem[\protect\citeauthoryear{Verbeek}{Verbeek}{2000}]%
        {verbeek2000daadkracht}
\bibfield{author}{\bibinfo{person}{Peter-Paul Camiel~Christiaan Verbeek}.}
  \bibinfo{year}{2000}\natexlab{}.
\newblock \bibinfo{booktitle}{\emph{De daadkracht der dingen: over techniek,
  filosofie en vormgeving}}.
\newblock \bibinfo{publisher}{Boom Koninklijke Uitgevers}.
\newblock


\bibitem[\protect\citeauthoryear{Whitby}{Whitby}{2008}]%
        {whitby2008sometimes}
\bibfield{author}{\bibinfo{person}{Blay Whitby}.}
  \bibinfo{year}{2008}\natexlab{}.
\newblock \showarticletitle{Sometimes it’s hard to be a robot: A call for
  action on the ethics of abusing artificial agents}.
\newblock \bibinfo{journal}{\emph{Interacting with Computers}}
  \bibinfo{volume}{20}, \bibinfo{number}{3} (\bibinfo{year}{2008}),
  \bibinfo{pages}{326--333}.
\newblock


\bibitem[\protect\citeauthoryear{Wilson, Hoffman, and Morgenstern}{Wilson
  et~al\mbox{.}}{2019}]%
        {wilson2019predictive}
\bibfield{author}{\bibinfo{person}{Benjamin Wilson}, \bibinfo{person}{Judy
  Hoffman}, {and} \bibinfo{person}{Jamie Morgenstern}.}
  \bibinfo{year}{2019}\natexlab{}.
\newblock \showarticletitle{Predictive inequity in object detection}.
\newblock \bibinfo{journal}{\emph{arXiv preprint arXiv:1902.11097}}
  (\bibinfo{year}{2019}).
\newblock


\bibitem[\protect\citeauthoryear{Zuboff}{Zuboff}{2019}]%
        {zuboff2019age}
\bibfield{author}{\bibinfo{person}{Shoshana Zuboff}.}
  \bibinfo{year}{2019}\natexlab{}.
\newblock \bibinfo{booktitle}{\emph{The age of surveillance capitalism: The
  fight for a human future at the new frontier of power}}.
\newblock \bibinfo{publisher}{Profile Books}.
\newblock


\end{thebibliography}

%

\end{document}